\documentstyle[12pt]{article}
\newcommand{\be}{\begin{equation}}
\newcommand{\ee}{\end{equation}}
\newcommand{\R}{\rm I \mkern -3mu R}
\newcommand{\N}{\rm I \mkern -3mu N}
\begin{document}
\thispagestyle{empty}
\begin{flushright}
IFA-FT-400-1994, October
\end{flushright}
\bigskip\bigskip\begin{center}
{\bf \Large{ON EQUATIONS WITH\\
{}~\\ UNIVERSAL INVARIANCE}}
\end{center}
\vskip 1.0truecm
\centerline{\bf
D. R. Grigore\footnote{e-mail: grigore@roifa.bitnet, grigore@ifa.ro}}
\vskip5mm
\centerline{Dept. Theor. Phys., Inst. Atomic Phys.,}
\centerline{Bucharest-M\u agurele, P. O. Box MG 6, ROM\^ANIA}
\vskip 2cm
\bigskip \nopagebreak \begin{abstract}
\noindent
A general discussion of equations with universal invariance for
a scalar field is provided in the framework of Lagrangian theory
of first-order systems.
\end{abstract}
\newpage\setcounter{page}1

\section{Introduction}

Recently there has been some interest in the study of the
partial differential equations having the so-called {\it
universal invariance} \cite{F1}-\cite{F4}. For a field with $N$
components this means that if
$
\Phi^{A}~(A=1,2,...,N)
$
is a solution of the field equations, then
$
F\circ\Phi
$
is also a solution of the same equation for any diffeomorphism
$
F \in Diff(\R^{N}).
$
One usually supposes that such an equation follows from a
variational principle i.e. is of the Lagrangian type.

The principle of universal invariance seems to produce many
interesting equations of physical relevance. So, it will be
desirable to have a program of classifying such equations
following from the characterization above. We will start in this
paper with the simplest case namely when the Lagrangian is of the
first order and the field is scalar i.e.
$N = 1.$

In the case of first order Lagrangians one can use the formalism
described in \cite{G} which is very well suited for the study of
Lagrangian systems with groups of symmetries. Applying this
formalism we will be able to write down a rather general
equation with universal invariance for a scalar field.

In Section 2 we present the general formalism for the case of a
scalar field. In this case a rather complete discussion is
possible. In Section 3 we derive the result announced above.

\section{A Geometric Setting for the Lagrangian Formalism}

2.1 The geometric setting of the Lagrangian theory in particle
mechanics is usually based on the Poincar\'e-Cartan 1-form, but
it is also possible to use a 2-form having as the associated
system exactly the Euler-Lagrange equations \cite{Kl}. This point
of view was intensively exploited by Souriau \cite{S} in
connection with the Hamiltonian formalism. The proper
generalization of these ideas to classical field theory is due
to Krupka, Betounes and Rund \cite{Kr}-\cite{R}. We will follow
the presentation from \cite{G}, but we will study, for symplicity,
directly the case of a sclar field.

2.2 Let $S$ be a differentiable manifold of dimension
$n+1$. The first order Lagrangian formalism is based on an
auxiliary object, namely the bundle of 1-jets of
$n$-dimensional submanifolds of $S$,
$$
J^{1}_{n}(S) \equiv \cup_{p \in S} J^{1}_{n}(S)_{p}
$$
where
$
J^{1}_{n}(S)_{p}
$
is the manifold of
$n$-dimensional linear subspaces of the tangent
space
$
T_{p}(S)
$ at $S$ in the point $p \in S$.
This manifold is naturally fibered over $S$;
let us denote by
$
\pi
$
the canonical projection
and construct a system of charts adapted to this fibered structure.
We choose a system of local coordinates
$
(x^{\mu},\psi)
$
on the open set
$
U \subseteq S;
$
here
$
\mu =1,...,n.
$
Then on the open set
$
V \subseteq \pi^{-1}(U)
$
we shall choose the local coordinate system
$
(x^{\mu},\psi,\psi_{\mu})
$
defined as follows: if
$
(x^{\mu},\psi)
$
are the coordinates of
$
p \in U
$
then the $n$-dimensional hyperplane in
$
T_{p}(S)
$
corresponding to
$
(x^{\mu},\psi,\psi_{\mu})
$
is spanned by the tangent vectors:
\be
{\delta \over \delta x^{\mu}} \equiv {\partial \over \partial x^{\mu}} +
\psi_{\mu} {\partial \over \partial \psi}.
\ee

We will systematically use the summation convention
over the dummy indices.

By an {\it evolution space} we mean any (open) subbundle
$E$
of
$
J^{1}_{n}(S).
$
Note that
$
dim(J^{1}_{n}(S)) = 2n+1.
$

2.3 Let us define for any evolution space $E$:
\be
\Lambda_{LS} \equiv \{ \sigma \in \wedge^{n+1}(J^{1}_{n}(S))|
i_{Z_{1}}i_{Z_{2}} \sigma=0,\forall Z_{1},Z_{2} \in Vect(E)~vertical\}.
\ee

A vector field
$
Z \in Vect(E)
$
is {\it vertical} if and only if
$
\pi_{*} Z = 0.
$
It is clear that any
$
\sigma \in \Lambda_{LS}
$
can be written in the local coordinates from above as follows:
\begin{eqnarray}
\sigma = \varepsilon_{\mu_{1},...,\mu_{n}}
(\sigma^{\mu_{0}} d\chi_{\mu_{0}} \wedge dx^{\mu_{1}}
\wedge...\wedge dx^{\mu_{n}} +
n\sigma^{\mu_{0}\mu_{1}}d\chi_{\mu_{0}} \wedge \delta\psi \wedge dx^{\mu_{2}}
\wedge...\wedge dx^{\mu_{n}}) \\ \nonumber
+ n! \tau \delta\psi \wedge dx^{1} \wedge...\wedge dx^{n}.
\end{eqnarray}

Here
$
\sigma^{\mu},
\sigma^{\mu_{0}\mu_{1}}
$
and
$
\tau
$
are smooth functions on
$E$,
\be
\delta\psi \equiv d\psi - \psi_{\mu} dx^{\mu}
\ee
and
$
\varepsilon_{\mu_{1},...,\mu_{n}}
$
is the signature of the permutation
$
(1,...,n) \mapsto (\mu_{1},...,\mu_{n}).
$

One can verify directly by performing a change of charts
$
(x^{\mu},\psi) \mapsto (y^{\mu},\zeta)
$
inducing
$
(x^{\mu},\psi,\psi_{\mu}) \mapsto (y^{\mu},\zeta,\zeta_{\mu})
$
that the following equations have an intrinsic global meaning:
\be
\sigma^{\mu} = 0
\ee
\be
\sigma^{\mu\nu} = \sigma^{\nu\mu} .
\ee

Any closed element
$
\sigma \in \Lambda_{LS}
$
verifying (5) and (6) will be called a
{\it Lagrange-Souriau form on E} (LS-form). Such a
$
\sigma
$
is of the form:
\be
\sigma = n \varepsilon_{\mu_{1},...,\mu_{n}}
\sigma^{\mu_{0}\mu_{1}} d\psi_{\mu_{0}}
\wedge \delta\psi \wedge dx^{\mu_{2}} \wedge...\wedge dx^{\mu_{n}} +
n!\tau \delta\psi \wedge dx^{1} \wedge...\wedge dx^{n}.
\ee

The closedness condition
\be
d \sigma = 0
\ee
gives explicitely:
\be
{\partial \sigma^{\mu\nu}\over \partial {\psi_{\rho}}} =
{\partial \sigma^{\mu\rho}\over \partial {\psi_{\nu}}}
\ee
\be
{\delta \sigma^{\mu\nu}\over \delta {x^{\nu}}} +
{\partial \tau\over \partial \psi_{\mu}} = 0.
\ee

We will call (6)+(9)+(10) the {\it structure equations}.

A {\it Lagrangian system over S} is a couple
$
(E,\sigma)
$
with
$
E \subseteq J^{1}_{n}(S)
$
an evolution space over
$S$
and
$
\sigma
$
a Lagrange-Souriau form on $E$.

There is a natural {\it equivalence} relation between two such systems,
$
(E_{1},\sigma_{1})
$
and
$
(E_{2},\sigma_{2})
$
over the same manifold
$S$
i.e. one must have a map
$
\alpha \in Diff(S)
$
such that
$
\dot{\alpha}(E_{1}) = E_{2}
$
and:
\be
(\dot\alpha)^{*}\sigma_{2} = \sigma_{1}.
\ee
where
$
\dot\alpha \in Diff(J^{1}_{n}(S))
$
is the natural lift of
$
\alpha.
$

2.4 An {\it evolutions} is an immersions
$
\Psi:M \rightarrow S,
$
where
$M$
is some $n$-dimensional manifold (the "space-time" manifold).

Let
$
(E,\sigma)
$
be a Lagrangian system over
$S$; one says that
$
\Psi:M \rightarrow S,
$
verifies the {\it Euler-Lagrange equations} if
\be
(\dot\Psi)^{*} i_{Z}\sigma=0
\ee
for any vector field
$
Z \in Vect(E).
$
Here
$
\dot\Psi:M \rightarrow J^{1}_{n}(S)
$
is the natural lift of
$
\Psi.
$

In local coordinates one can arrange such that
$
\Psi
$
has the form
$
x^{\mu} \mapsto (x^{\mu},\Psi(x));
$
then
$
\dot\Psi:M \rightarrow J^{1}_{n}(S)
$
is given by
$
x^{\mu} \mapsto (x^{\mu},\Psi(x),{\partial \Psi\over \partial x^{\mu}}(x))
$
and (12) have the local expression:
\be
\sigma^{\mu\nu} \circ \dot\Psi {\partial^{2} \Psi \over \partial
x^{\mu}\partial x^{\nu}} - \tau\circ \dot\Psi = 0.
\ee

An interesting result following directly from this equation is

{\bf Lemma} The Euler-Lagrange equations  are trivial
{\it iff} $\sigma = 0$.

2.5 We come now to the notion of symmetry. By a {\it symmetry} of
the Euler-Lagrange equations we understand a map
$
\phi \in Diff(S)
$
such that if
$
\Psi:M \rightarrow S
$
is a solution of these equations, then
$
\phi \circ \Psi
$
is a solution of these equations also.

In the case of a scalar field one can completely describe the
structure of a symmetry. We have:

{\bf Theorem 1:} {\it Let
$
(E,\sigma)
$
be a Lagrangian system for a scalar field and
$
\phi \in Diff(S)
$
a symmetry. Then there exists
$
\rho \in {\cal F}(E)
$
such that
\be
(\dot\phi)^{*} \sigma= \rho \sigma.
\ee

The function
$
\rho
$
must satisfy the equation
\be
d\rho \wedge \sigma = 0
\ee
or, in local coordinates:
\be
\tau {\partial \rho\over \partial \psi_{\mu}} +
\sigma^{\mu\nu} {\delta \rho \over\delta x^{\nu}} = 0
\ee
\be
\sigma^{\mu\nu} {\partial \rho \over\partial  \psi_{\lambda}} -
\sigma^{\mu\lambda} {\partial \rho \over\partial \psi_{\nu}} = 0.
\ee}

{\bf Proof:} Because
$Z$
in (12) is arbitrary, one easily descovers that
$
\phi
$
is a symmetry {\it iff}
$$
(\dot\Psi)^{*} i_{Z}\sigma=0 \Longrightarrow
(\dot\Psi)^{*} (\dot\phi)^{*} i_{Z}\sigma=0,~\forall Z \in Vect(E),~
\forall \Psi:M \mapsto S.
$$

We denote for simplicity
\be
\tilde{\sigma} \equiv (\dot\phi)^{*} \sigma.
\ee

One can show immediately that
$
\tilde{\sigma}
$
is a LS-form so it has the structure given by (7) with
$
\sigma^{\mu\nu} \mapsto \widetilde{\sigma^{\mu\nu}}
$
and
$
\tau \mapsto \tilde{\tau}.
$
It follows from above that we have:
\be
\sigma^{\mu\nu} \circ \dot\Psi {\partial^{2} \Psi \over \partial
x^{\mu}\partial x^{\nu}} - \tau\circ \dot\Psi = 0 \Longrightarrow
\widetilde{\sigma^{\mu\nu}} \circ \dot\Psi {\partial^{2} \Psi \over \partial
x^{\mu}\partial x^{\nu}} - \tilde{\tau}\circ \dot\Psi = 0
\ee
(see (13).) Equivalently:
\be
\sigma^{\mu\nu} \psi_{\{\mu\nu\}} - \tau = 0 \Longrightarrow
\widetilde{\sigma^{\{\mu\nu\}}} \psi_{\{\mu\nu\}} - \tilde{\tau} = 0
\ee
where
$
\psi_{\{\mu\nu\}}
$
is an arbitrary real symmetric matrix. In fact, it is more
appropriate to consider expressions of the type appearing in
(20) as functions on
$
J^{2}_{n}(S).
$

It is not hard to prove that (20) implies the existence of
$
\rho \in {\cal F}(E)
$
such that:
$$
\widetilde{\sigma^{\mu\nu}} \psi_{\{\mu\nu\}} - \tilde{\tau} =
\rho (\sigma^{\mu\nu} \psi_{\{\mu\nu\}} - \tau) \Longleftrightarrow
$$
$$
\tilde{\tau} = \rho \tau~~,
\widetilde{\sigma^{\mu\nu}} = \rho \sigma^{\mu\nu}.
$$

So, we find out that:
\be
\tilde{\sigma} = \rho \sigma.
\ee

But, as noted above,
$
\tilde{\sigma}
$
is a LS-form, so
$
\rho\sigma
$
must be a LS-form. From the definition of a LS-form it is clear
that only the closedness condition (15) is missing. The
derivation of (16) and (17) is elementary.
$
\Box
$

{\bf Remarks:}

1) If
$
\rho
$
is not locally constant, then (15) implies that
$
\sigma
$
is of the form
\be
\sigma = d\rho \wedge \omega
\ee
with
$
\omega
$
a $n$-form.

2) Let us suppose that the Lagrangian system
$
(E,\sigma)
$
is {\it non-degenerated}, that's it:
\be
det(\sigma^{\mu\nu}) \not= 0.
\ee
This condition has a global intrinsec meaning as it easily
follows performing a change of charts. (The condition of
non-degeneracy ensures that the Euler-Lagrange equations (13)
can be "solved" with respect to the second order derivatives and
the Cauchy problem can be well defined.) If we have (23) then
one finds from (17) that
$
{\partial \rho\over \partial \psi_{\lambda}} = 0.
$
Next, (16) gives
$
{\partial \rho\over \partial \psi} = 0
$
and
$
{\partial \rho\over \partial x^{\mu}} = 0
$
so
$
\rho
$
is locally constant. This result is a sort of Lee-Hwa Chung
theorem (see e.g. \cite{SM}) for the Lagrangian formalism.

3) The case
$
\rho = 1
$
corresponds to the so-called {\it Noetherian symmetries}. For a
detailed discussion see \cite{G}.
$
\Box
$

If a group
$G$
act on
$S$:
$
G \ni g \mapsto \phi_{g} \in Diff(S)
$
then we say that
$G$
is a {\it group of (Noetherian) symmetries} for
$
(E,\sigma)
$
if for any
$
g \in G,
$
$
\phi_{g}
$
is a (Noetherian) symmetry. In particular we have:
\be
(\dot\phi_{g})^{*}\sigma=\rho_{g} \sigma.
\ee

It is considered of physical interest to solve the
following classification problem: given the manifold
$S$
with an action of some group
$G$
on
$S,$
find all Lagrangian systems
$
(E,\sigma)
$
where
$
E \subseteq J^{1}_{n}(S)
$ is on open subset and
$G$
is a group of (Noetherian) symmetries for
$
(E,\sigma).
$
This goal will be achieved by solving simultaneously
(6), (9), (10), (16), (17) and (24) in local coordinates and
then investigating the possibility of globalizing
the result.

2.6 We close this section explaining the connection with the usual
Lagrangian formalism. We can consider that the
open set $V \subseteq \pi^{-1}(U)$ is simply
connected by choosing it small enough.

{}From the structure equations (6), (9) and (10) one easily finds
out that there exists a (local) function
$L$
on
$V$
such that:
\be
\sigma^{\mu\nu} ={\partial^{2} L \over \partial\psi_{\mu}
\partial\psi_{\nu}}
\ee
and
\be
\tau = {\partial L\over \partial \psi} - {\delta \over \delta
x^{\mu}} \left({\partial L\over \partial \psi_{\mu}}\right).
\ee

Now (13) takes the usual form for the Euler-Lagrange equations.
$L$
is called a {\it local Lagrangian}. If
$
\sigma
$
is given by (7) but with the coefficient functions as in (25)
and (26) above, then we denote it by
$
\sigma_{L}.
$

\section{Universal Invariance}

3.1 In the framework developped in Section 2, let us consider that
$
S = \R^{n} \times \R
$
with global coordinates
$
(x^{\mu},\psi),~(\mu =1,...,n).
$
We can take
$
E = J^{1}_{n}(S) \simeq \R^{n} \times \R \times \R^{n}
$
with global coordinates
$
(x^{\mu},\psi,\psi_{\mu}).
$
If
$F \in Diff(\R),
$
let us consider
$
\phi_{F} \in Diff(S)
$
given by:
\be
\phi_{F}(x,\psi) = (x,F(\psi)).
\ee

By definition, the Lagrangian system
$
(E,\sigma)
$
defined above has {\it universal invariance} if
$
\phi_{F}
$
is a symmetry, i.e. (see (24)) we have:
\be
(\dot\phi_{F})^{*} \sigma = \rho_{F}~\sigma
\ee
for some functions
$
\rho_{F} \in {\cal F}(E)
$
verifying (16) and (17). It is easy to show that
$
\rho_{F}
$
must verify the following consistency relation:
\be
\rho_{F_{1}} \circ \dot\phi_{F_{2}}~\rho_{F_{2}} =
\rho_{F{1}\circ F_{2}}
\ee
which is easily recognized as a cocycle condition. One would be
tempted to try to solve this cohomology problem. This can be
done under some reasonable smoothness conditions, but we will
prefer to circumvent this analysis.

3.2 We substitute (7) into (28) and get equivalenty:
\be
(F')^{2} \sigma^{\mu\nu} \circ \dot\phi_{F} = \rho_{F}~
\sigma^{\mu\nu}
\ee
\be
F'(\tau \circ \dot\phi_{F} - F''\sigma^{\mu\nu}\circ
\dot\phi_{F}~\psi_{\mu} \psi_{\nu}) = \rho_{F}~\tau.
\ee

{}From (30) it easily follows that
$
\rho_{F}
$
is a coboundary i.e. is of the form
\be
\rho_{F} = b \circ \dot\phi_{F}~b^{-1}
\ee
for some function
$
b \in {\cal F}(E).
$
In fact, one can show that the most general solution of (29) is
of this type.

3.3 We will study in fact only a particular case which covers
the equations from \cite{F1}-\cite{F4}, namely when
$b$
is a homogeneous function of degree
$
p \in \N
$
in the variables
$
\psi_{\mu}.
$
Then from (32) we have:
\be
\rho_{F} = (F')^{p}.
\ee

We insert (33) into (30)+(31) and consider that
$F$
is an infinitesimal diffeomorphism i.e.
\be
F(\psi) = \psi + \theta(\psi)
\ee
with
$
\theta
$
infinitesimal but otherwise arbitrary. We obtain:
\be
{\partial \sigma^{\mu\nu} \over \partial\psi} = 0
\ee
\be
\psi_{\lambda} {\partial \sigma^{\mu\nu} \over \partial\psi_{\lambda}} =
(p - 2) \sigma^{\mu\nu}
\ee
\be
{\partial \tau \over \partial\psi} = 0
\ee
\be
\psi_{\lambda} {\partial \tau \over \partial\psi_{\lambda}} =
(p - 1) \tau
\ee
\be
\sigma^{\mu\nu} \psi_{\mu} \psi_{\nu} = 0.
\ee

{}From the consistency equation (16) we obtain
\be
\sigma^{\mu\nu} \psi_{\mu} = 0
\ee
so (39) is redundant. Equation (17) is identically satisfied.

3.4 We analyse now the system (35)-(38)+(40). First, we
concentrate on the functions
$
\sigma^{\mu\nu}.
$
Let us note that (36) is the infinitesimal form of the
homogeneity property:
\be
\sigma^{\mu\nu}(x,\lambda\psi_{\mu}) = \lambda^{p-2}
\sigma^{\mu\nu}(x,\psi_{\mu})~,\forall \lambda \in \R^{*}.
\ee

In the chart
$
\psi_{0} \not= 0
$
this means that
$
\sigma^{\mu\nu}
$
is of the following form:
\be
\sigma^{\mu\nu} = \psi_{0}^{p-2} s^{\mu\nu}\circ \pi
\ee
where
$
s^{\mu\nu}
$
is a smooth function of the variables
$
x, y_{1},...,y_{n-1}
$
and we have defined
\be
\pi(x, \psi_{1},...,\psi_{n}) = \left(x,{\psi_{1}\over\psi_{0}},...,
{\psi_{n-1}\over\psi_{0}}\right).
\ee

{}From (6) we have
\be
s^{\mu\nu} = s^{\nu\mu}
\ee
and from (40):
\be
s^{00} = \sum_{i,j=1}^{n-1} y_{i} y_{j} s^{ij}
\ee
\be
s^{0i} = - \sum_{j=1}^{n-1} y_{j} s^{ij}.
\ee

We still have at our disposal the structure equation (9). It is
convenient to define the operator
\be
D \equiv \sum_{j=1}^{n-1} y_{j} {\partial \over \partial y_{j}}.
\ee

Then (9) is equivalent to:
\be
{\partial s^{00} \over \partial y_{i}} = (p - 2 - D) s^{0i}
\ee
\be
{\partial s^{0i} \over \partial y_{i}} = (p - 2 - D) s^{ij}
\ee
\be
{\partial s^{ij} \over \partial y_{k}} =
{\partial s^{ik} \over \partial y_{j}}
\ee

If we insert (46) into (49) we obtain:
\be
(p - 1) s^{ij} = 0.
\ee

Analogously, if we insert (45) and (46) into (48) we get:
\be
(p - 1) \sum_{j=1}^{n-1} y_{j} s^{ij} = 0.
\ee

If
$
p \not= 1
$
then it easily follows that
\be
\sigma = 0
\ee
so we are left only with the case
$
p = 1.
$
In this case (51) and (52) are becoming identities and
$
s^{ij}
$
are constrained only by (44) and (50). It follows that there
exists a function
$l$
depending on
$x$
and
$
y_{1},...,y_{n-1}
$
such that
\be
s^{ij} = {\partial^{2} l\over \partial y_{i} \partial y_{j}}.
\ee

Then we get from (45) and (46):
\be
s^{i0} = - D {\partial l\over \partial y_{i}}
\ee
and
\be
s^{00} = (D^{2} - D)l.
\ee

The structure of the functions
$
s^{\mu\nu}
$
is completely elucidated. If we define:
\be
L_{0} \equiv \psi_{0} l\circ \pi
\ee
then it is elementary to prove that we have (25) with
$
L \rightarrow L_{0}.
$

If we define
\be
\sigma' = \sigma - \sigma_{L_{0}}
\ee
then it is easy to analyse the structure of this
auxilliary LS-form which also verifies the invariance condition
(28)+(33).

The final result can be summarized as follows:

{\bf Theorem 2:} {\it Let
$
(E,\sigma)
$
be a first-order Lagrangian system for a scalar field having
universal invariance (28) with
$
\rho_{F}
$
given by (33). Then we have non-trivial solutions only for
$
p = 1.
$
In this case we have
\be
\sigma = \sigma_{L}
\ee
with
\be
L = \psi_{0} l\circ \pi + \psi\tau.
\ee

Here
$l$
is a smooth function depending on
$x$
and
$
y_{1},...,y_{n-1}
$
and
$
\tau
$
is only
$x$-dependent.
The corresponding Euler-Lagrange equations (13) are, in the notations:
$
\Psi_{\mu} \equiv {\partial \Psi \over \partial x^{\mu}},~~
\Psi_{\{\mu\nu\}} \equiv {\partial^{2} \Psi \over
\partial x^{\mu}\partial x^{\nu}}
$
\be
\sum_{i,j=1}^{n-1} {\partial^{2} l\over \partial y_{i}\partial y_{j}}
\left(x,{\Psi_{i}\over\Psi_{0}}\right) \Psi_{0}^{-3} (\Psi_{i}
\Psi_{j}\Psi_{\{00\}} - \Psi_{0}\Psi_{i}\Psi_{\{0j\}} -
\Psi_{0}\Psi_{j}\Psi_{\{0i\}} + \Psi_{0}^{2}\Psi_{\{ij\}}) -
\tau = 0.
\ee}

{\bf Remarks:}

4) For
$
n = 2
$
we get
\be
{\partial^{2} l\over \partial y^{2}}\left(x,{\Psi_{1}\over\Psi_{0}}\right)
\Psi_{0}^{-3} (\Psi_{1}^{2}\Psi_{\{00\}} - 2\Psi_{0}\Psi_{1}\Psi_{\{01\}}
+ \Psi_{0}^{2}\Psi_{\{11\}}) - \tau = 0.
\ee

If we take
$
\tau = 0
$
we get
$
(\cdots) = 0
$
which is the equation appearing in \cite{F4}.

5) Because
$
p = 1,
$
the universal invariance is not a Noetherian symmetry.
$
\Box
$

\section{Final Comments}

There are a number of results obtained in this paper which will
interesting to generalize.

First, one could try to extend the analysis above to the case of
a field with more than one component. This extension seems
possible and plausible, but there might appear some technical
problems.

Next, we come to the universal invariance. Can one study the
general case (32)? This seems to be a rather complicated problem.

Finally, one should like to generalize these results to
higher-order Lagrangian systems. This problem is more manageable
and some results in this direction will be reported soon. We
will have to use a completely different method, because a
generalization of the formalism from section 2 to higher-order
Lagrangian systems is not available.


\begin{thebibliography}{99}
\bibitem{F1}
D. B. Fairlie, J. Govaerts, A. Morozov,
Nucl. Phys. B 373 (1992) 214-232

\bibitem{F2}
D. B. Fairlie, J. Govaerts,
Phys. Lett. B 281 (1992) 49-53

\bibitem{F3}
D. B. Fairlie, J. Govaerts,
Jour. Math. Phys. 33 (1992) 3543-3556

\bibitem{F4}
D. B. Fairlie, J. Govaerts,
"Linearization of Universal Field Equations",
preprint Durham DPT-92/47

\bibitem{G}
D. R. Grigore,
Fortschr. der Phys. 41 (1993) 569-617

\bibitem{Kl}
J. Klein,
Ann. Inst. Fourier (Grenoble) 12 (1962) 1-124

\bibitem{S}
J.M.Souriau,
"{\it Structure des Syst\`emes Dynamiques}",
Dunod, Paris, 1970
\bibitem{Kr}
D. Krupka,
Czech Math. Journ. 27 (1977) 114-118
\bibitem{B1}
D.Betounes,
Phys. Rev. D 29 (1984) 599-606
\bibitem{B2}
D.Betounes,
J. Math. Phys. 28 (1987) 2347-2353
\bibitem{R}
H.Rund,
Lect. Notes in Pure and Appl. Math. 100 (1985) 455-469
\bibitem{SM}
E. C. G. Sudarshan, N. Mukunda,
"{\it Classical Dynamics: A Modern Perspective},
Wiley, New York, 1974
\end{thebibliography}
\end{document}